\documentclass[pra,aps,twocolumn,longbibliography]{revtex4-2}

\usepackage[utf8]{inputenc}

\usepackage{xcolor}
\usepackage{amsmath}
\usepackage{amsfonts}
\usepackage{amssymb}
\usepackage{makeidx}
\usepackage{graphicx}
\usepackage{tikz}
\usepackage[normalem]{ulem}
\usepackage{float}
\usepackage{braket}

\usepackage[caption=false]{subfig}

\usepackage{color}
\usepackage[colorlinks=true,citecolor=blue,linkcolor=blue,urlcolor=blue]{hyperref}
\usepackage{url}

\DeclareMathOperator{\tr}{\text{tr}}

\begin{document}

\title{Squeezing light to get nonclassical work in quantum engines}
\author{A. Tejero}
\email{atejero@onsager.ugr.es}
\affiliation{Departamento de Electromagnetismo y F\'{\i}sica de la Materia, and Institute Carlos I for Theoretical and Computational Physics, University of Granada, 18071 Granada, Spain.}
\author{D. Manzano}
\email{manzano@onsager.ugr.es}
\affiliation{Departamento de Electromagnetismo y F\'{\i}sica de la Materia, and Institute Carlos I for Theoretical and Computational Physics, University of Granada, 18071 Granada, Spain.}
\author{P.I. Hurtado}
\email{phurtado@onsager.ugr.es}
\affiliation{Departamento de Electromagnetismo y F\'{\i}sica de la Materia, and Institute Carlos I for Theoretical and Computational Physics, University of Granada, 18071 Granada, Spain.}
\date{\today}

\begin{abstract}
Light can be \emph{squeezed} by reducing the quantum uncertainty of the electric field for some phases. We show how to use this purely quantum effect to extract net mechanical work from radiation pressure in a simple quantum photon engine. Along the way, we demonstrate that the standard definition of work in quantum systems does not capture the extractable mechanical work, as it does not reflect the energy leaked to these quantum degrees of freedom. We use these results to design an Otto engine able to produce mechanical work from squeezing baths, in the absence of a thermal gradient. Interestingly, while work extraction from squeezing generally improves for low temperatures, there exists a nontrivial squeezing-dependent temperature for which work production is maximal, demonstrating the complex interplay between thermal and squeezing effects.
\end{abstract}
\maketitle

\section{Introduction}
In quantum heat engines \cite{scovil59a,alicki:jpa79}, heat is converted into work using a quantum system as working medium. Interest in these engines has grown steadily in recent years due to their potential quantum advantages over classical engines \cite{scully:science03, linden:prl10,  scully:pnas11, rossnagel14a, hardal15a, uzdin:prx15, roulet:pre17, seah:njp18, zheng:pre16, binder_18, bhattacharjee:epjb21, rossnagel:science16, maslennikov19a, lindenfels19a, klatzow19a, ryan08a, peterson19a, ono20a, koski14a,cangemi:physrep24}, also driving research in quantum thermodynamics \cite{vinjanampathy:cp16}, with the aim of efficiently converting heat into work in microscopic systems subject to both thermal and quantum fluctuations. Recent technological advances have led to different experimental realizations of these engines \cite{rossnagel:science16, maslennikov19a, lindenfels19a, klatzow19a, ryan08a, peterson19a, ono20a, koski14a, martinez:sm17, bhattacharjee:epjb21}, revealing net quantum advantages in many cases. Most of these efforts have been based on a mathematical definition of quantum work first proposed by Alicki \cite{alicki:jpa79}, which has, however, been challenged in several cases \cite{weimer08, seegebrecht:jsp2024,ahmadi:scirep2023,alipour:pra2022,bera:natcomm2017}. A significant limitation of Alicki’s work is its inability to account for internal energy conversions within the system. As a result, it does not provide a complete prediction of the mechanical work extractable from a given process \cite{weimer08}. An alternative definition, based on the expansion work exerted by radiation pressure in optomechanical systems, has been recently put forward \cite{tejero:pre24}, allowing to address this fundamental issue in a well-defined manner.

A key question is whether net work can be efficiently extracted from purely quantum degrees of freedom, thus departing from the classical description. One of these quantum properties of light is \emph{squeezing} \cite{walls83a}, i.e., the possibility that the electric-field strength for some particular phase has a smaller quantum uncertainty than the corresponding uncertainty of a coherent light state \cite{loudon:jmo1987, gerry_04, walls_2025, garrison_08}. In particular, for an electromagnetic wave of phase $\phi$, the (dimensionless) amplitude of its electric field is given by the quadrature operator 
\begin{equation}
X_{\phi}=\frac{1}{\sqrt{2}}\left(\textrm{e}^{-i\phi} a + \textrm{e}^{i \phi} a^\dagger  \right)=\cos(\phi) X + \sin(\phi) Y, 
\end{equation}
with $a^\dagger~(a)$ the creation (destruction) operators for the harmonic oscillator, while operators $X$ and $Y$ represent the electric-field strength at $\phi=0$ and $\phi=\pi/2$, respectively. These quadrature operators fulfill an uncertainty relation in the form $\sigma^2_X \sigma^2_Y \ge 1/16$, with $\sigma^2_A$ the variance associated with a generic operator $A$. In the vacuum state $\ket{0}$, this relation is saturated and $\sigma^2_X=\sigma^2_Y=1/4$. In contrast, a state of light is squeezed if there is a phase $\phi$ such that $\sigma_{X_{\phi}}<1/4$, meaning that there is less uncertainty in one quadrature component at the price of having more uncertainty in another (antisqueezed) component \cite{walls83a,loudon:jmo1987, gerry_04, walls_2025, garrison_08}. Squeezed states of light belong to the so-called \emph{nonclassical states of light}, and they have found important applications in experimental physics, e.g., to increase the detection rate of gravitational waves at LIGO \cite{tse19a,acernese19a,abbott21a,ligo:science24}. More generally, the idea of using quantum effects such as squeezing as thermodynamic resources \cite{brandao:prl2013, binder_18,goold:jpa2016} has received considerable attention in the last years, for instance to generate work beyond classical limits \cite{scully:science03,dillenschneider:epl2009, horodecki:natcomm2013,perernau:prx2015,brown:njp2016,huang:pre2012, niedenzu:natcomm2018,manzano_pre16} or to charge quantum batteries \cite{centrone:2021pra}. However, these approaches mix thermodynamic and quantum drivings, making the analysis of the possible quantum advantages more complicated.

In this work, we use squeezing as a quantum resource to generate mechanical work from light via the radiation pressure exerted on an optomechanical mirror \cite{tejero:pre24}. Our analysis demonstrates that Alicki's definition of work in quantum systems, based on the expected time variation of the Hamiltonian operator \cite{alicki:jpa79}, does not capture the extractable mechanical work in this context. The reason is that Alicki's definition does not discriminate the energy leaked to maintain squeezing via two-photon processes. We show this difference both analytically for small light squeezing, and numerically for arbitrary squeezing parameters. 
We then use these results to design a quantum Otto engine to produce net mechanical work based on the injection and extraction of squeezing, in the absence of any thermal gradient, decoupling in this way thermodynamic and quantum driving forces. Heat exchanges with the squeezed baths and degrees of freedom are also monitored, establishing the thermodynamic consistency of the cycle. Interestingly, we show that while work extraction from squeezing generally improves for low temperatures, there exists a nontrivial squeezing-dependent temperature for which work production is maximal. This demonstrates the complex interplay between thermal and squeezing effects. 

\section{Master equation with squeezing baths}
\begin{figure}[t]
\centering
\includegraphics[width=0.95\linewidth]{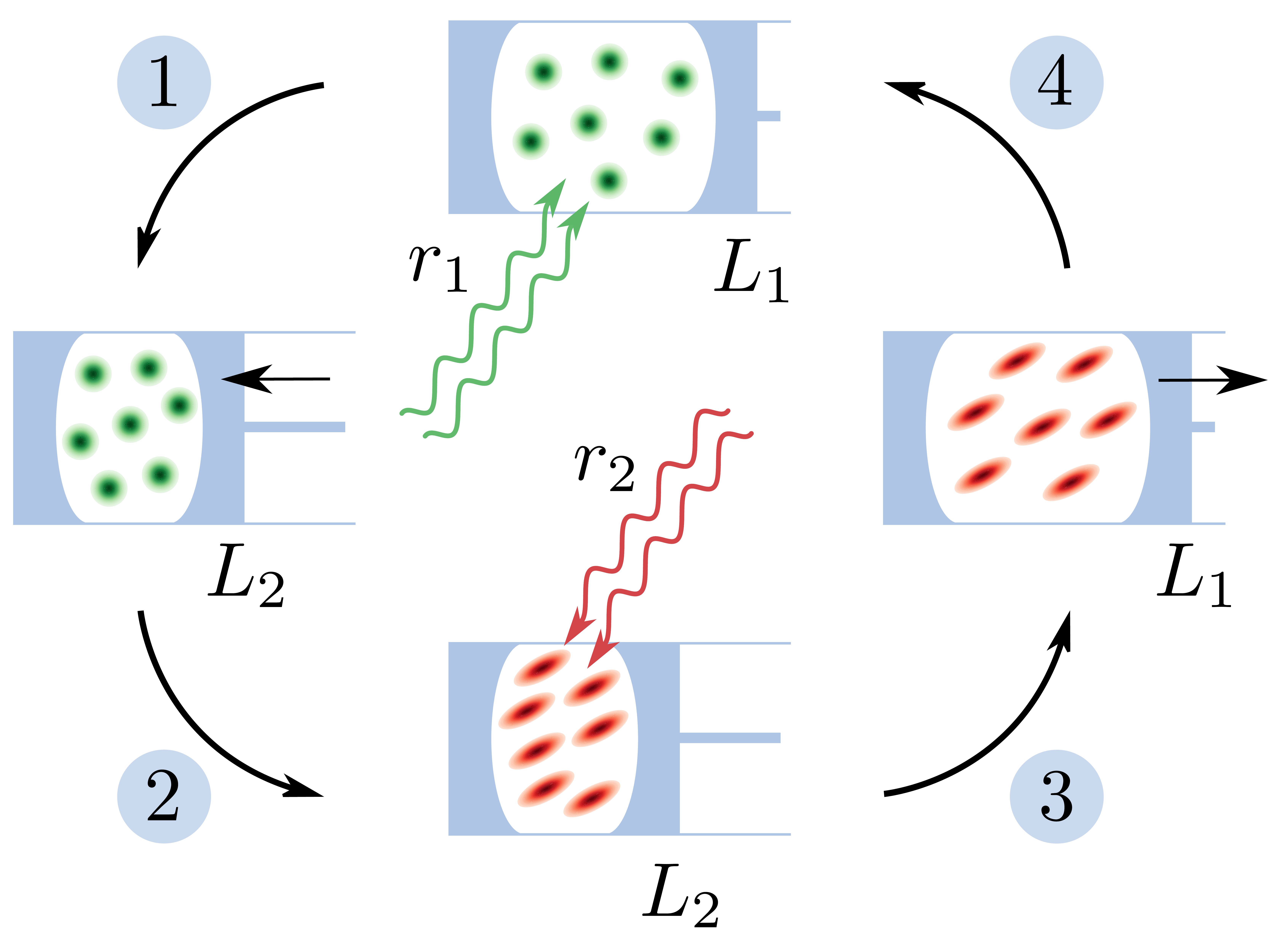}
\vspace{-0.5cm}
\caption{Sketch of the squeezing Otto engine. A single-mode cavity of section ${\cal S}$ and length $L_1$ is initially prepared in a squeezed thermal state with average photon number $\bar{n}$ and squeezing amplitude $r_1$ (top panel). The cavity is first compressed to length $L_2<L_1$ (stroke 1), then squeezed to $r_2\ne r_1$ (stroke 2). Next, we expand back to $L_1$ at constant squeezing $r_2$ (stroke 3), and it is finally brought back to the initial state by contact with a squeezing bath at $r_1$ (stroke 4).}
\label{fig:sketch}
\end{figure}

Our model is based on a single-mode cavity with a mobile mirror \cite{tejero:pre24}, see Fig.~\ref{fig:sketch}. The Hamiltonian of the system is 
\begin{equation}
H(t)= \hbar \omega(t) a^\dagger a, 
\label{hamilt}
\end{equation}
with $\omega(t)$ the natural frequency of the cavity, which may depend on time as it is inversely proportional to its length, $\omega(t)=c/L(t)$, with $c=\omega(0) L(0)$ a constant. Single-mode general squeezed pure states \cite{walls83a,loudon:jmo1987, gerry_04, walls_2025, garrison_08} are defined by two complex parameters $\alpha,\xi \in \mathbb{C}$ as $\Ket{\psi} = \Ket{\alpha,\xi}=S(\xi)D(\alpha)\Ket{0}$, where
\begin{equation}
\begin{split}
S(\xi) &= \exp[(\xi^* a^2 - \xi a^{\dagger^2})/2],\\
D(\alpha) &= \exp(\alpha a^\dagger - \alpha^* a),
\end{split}
\label{eq:sqop}
\end{equation}
are the squeezing and displacement operators, respectively. In particular, the parameter $\xi=r e^{i\varphi}$ will be used here to determine the squeezing of our system. Moreover, in many cases we will describe the squeezing state of a system only by the amplitude $r$, implying $\varphi=0$ for these cases. 

In an open quantum setting, squeezing can be pumped into the system by coupling the mode to a squeezed bosonic heat bath \cite{lu:pra1989,ekert:pra1990}. This can be modeled by a Lindblad-type equation for the density matrix $\rho(t)$ \cite{manzano:aip20,rivas_12,gardiner_00} (see Appendix \ref{App1})
\begin{eqnarray}
\dfrac{d \rho(t)}{dt} &=& - \dfrac{i}{\hbar} \left[H(t),\rho(t) \right] + \gamma (\bar{n} + 1) \Big(\hat{a}\rho(t) \hat{a}^{\dagger} - \label{eqn:lindblad_squeezing} \\ 
&-& \dfrac{1}{2} \left\{\hat{a}^{\dagger} \hat{a}, \rho(t) \right\}\Big) + \gamma \bar{n} \Big(\hat{a}^{\dagger}\rho(t) \hat{a} - \dfrac{1}{2} \left\{\hat{a} \hat{a}^{\dagger}, \rho(t) \right\}\Big), \nonumber
\end{eqnarray}
where $\gamma$ is the coupling strength to the bath. The average number of photons in the bath in resonance with the cavity is given by the Bose-Einstein distribution $\bar{n} = \left(\exp\left[\beta \hbar \omega\right] -1\right)^{-1}$, with 
$\beta=1/(k_{\mathrm{B}} T)$ the inverse temperature. Moreover, a \emph{squeezed} operator is defined as $\hat{A}\equiv S(\xi)A S(\xi)^{\dagger} = A \cosh\xi + A^\dagger \sinh \xi$. It can be shown that a system interacting with a squeezed bosonic bath with inverse temperature $\beta$ and squeezing parameter $\xi$ will eventually reach a squeezed thermal state  \cite{lu:pra1989,ekert:pra1990} 
\begin{equation}
\rho_\beta^{\xi}(t) = \frac{1}{\mathcal{Z}} S(\xi) \textrm{e}^{-\beta H(t)} S^{\dagger}(\xi), 
\label{eq:squeezed}
\end{equation}
with $\mathcal{Z}^{-1}=1-\textrm{e}^{-\beta\hbar\omega(t)}$ the unsqueezed partition function [as $S(\xi)$ is unitary].

\section{Extractable mechanical work}

In open quantum systems, the rate of change of the average energy can always be decomposed as 
\begin{equation}
\langle \dot{E}(t) \rangle= \tr[\dot{H}(t) \rho(t)] + \tr \left[ H(t) \dot{\rho}(t)\right], 
\end{equation}
where the dot denotes the derivative with respect to time.
Identifying quantum work with variations in energy due to changes in the system Hamiltonian, as first done by Alicki \cite{alicki:jpa79}, one thus recovers the customary definition of work in quantum thermodynamics, 
\begin{equation}
W_{\text{Al}} \equiv - \int_{t_0}^{t_f} \tr \left[\dot{H}(t) \rho(t)\right] dt,
\end{equation}
where $t_0$ and $t_f$ are the initial and final times for a given process. For our cavity, see Eq.~\eqref{hamilt}, Alicki's definition thus leads to  
\begin{equation}
W_{\text{Al}} = -\hbar  \int_{t_0}^{t_f} \dot{\omega}(t) \tr [a^\dagger a \rho(t)] dt = -\hbar  \int_{t_0}^{t_f} \dot{\omega}(t) \langle n(t)\rangle dt , 
\end{equation}
with $\langle n(t)\rangle$ the average number of photons in state $\rho(t)$. 

This definition of work, though widely used, has been challenged in several cases \cite{weimer08, seegebrecht:jsp2024, ahmadi:scirep2023, alipour:pra2022, bera:natcomm2017}. Moreover, a well-known significant limitation of $W_\text{Al}$ is its inability to account for internal-energy conversions within the system of interest. As a result, it does not always provide a precise prediction of the mechanical work that can be extracted from a given process \cite{weimer08}. To calculate the extractable work, an alternative definition has recently been proposed in the context of optomechanical machines, based on the expansion work driven by radiation pressure \cite{tejero:pre24}. In particular, allowing a mobile mirror in an optomechanical system to displace an infinitesimal length $\delta x$, the expansion work performed can be written as $\delta W= F \delta x$, where the force $F$ is exerted by the radiation pressure in the cavity. In this way, the average expansion work when a cavity of section ${\cal S}$ and length $L(t)$ changes volume from $V_0={\cal S} L(t_0)$ to $V_f={\cal S} L(t_f)$ is 
\begin{equation}
W_{\text{exp}}=\int_{V_0}^{V_f}  \braket{p(t)} dV = \int_{t_0}^{t_f}   \tr \left[\pi(t) \rho(t) \right]  {\cal S} \dot{L}(t) dt ,
\label{eqn:expansion}
\end{equation}
where $\braket{p(t)}=\tr \left[\pi(t) \rho(t) \right]$ is the average pressure in the cavity for state $\rho(t)$, with 
\begin{equation}
\pi (t) = \kappa(t) \left( a^\dagger a + a a^\dagger -aa\textrm{e}^{-i2\omega(t) t} -a^\dagger a^\dagger\textrm{e}^{i2\omega(t) t}  \right) 
\end{equation}
the radiation pressure operator, and $\kappa(t) = \hbar \omega(t)/[2 {\cal S} L(t)]$. Using that $[a,a^\dagger]=1$ and $L(t)=c/\omega(t)$ so $\dot{L}(t)=-c \dot{\omega}(t)/\omega(t)^2$, we hence can relate Alicki's and expansion works as $\tilde{W}_{\text{Al}} = W_{\text{exp}} +  \Delta W$, with
\begin{equation}
    \tilde{W}_{\text{Al}}\equiv -\hbar  \int_{t_0}^{t_f} \dot{\omega}(t) \tr \left[\left(a^\dagger a + \frac{1}{2}\right)\rho(t)\right] dt,
\end{equation}
Alicki's work including the zero-point energy (which can be renormalized out), and
\begin{equation}
 \Delta W \equiv - \frac{\hbar}{2} \int_{t_0}^{t_f} \dot{\omega}(t) \tr \left[\left(a^2 \textrm{e}^{-i2t \omega} + {a^\dagger}^2 \textrm{e}^{i2t \omega}\right)\rho(t)\right] dt ,
\label{DeltaW}
\end{equation}
which measures the energy leaked to the internal degrees of freedom of the cavity via two-photon processes. The two definitions of quantum work, $W_{\text{Al}}$ and $W_{\text{exp}}$, have been shown to be equivalent in an expanding cavity with no squeezing, analytically for quasistatic transformations and numerically for finite-time operations \cite{tejero:pre24}. 

To study the effect of nonclassical states of light, we consider now the \emph{quasistatic} expansion of a cavity in local equilibrium with a bath at inverse temperature $\beta$ and squeezing parameter $\xi$. As the process is quasistatic, we assume the cavity state at any moment to be a thermal squeezed state $\rho(t)=\rho_\beta^{\xi}(t)$, see Eq.~(\ref{eq:squeezed}). Crucially, the squeezing operator $S(\xi)= \exp[(\xi^* a^2 - \xi a^{\dagger^2})/2]$ connects Fock states $\ket{n}$ and $\ket{n\pm2}$ via two-photon processes associated to operators $a^2$ and ${a^\dagger}^2$. These two-photon excitation/decay processes will lead in general,  and as far as the squeezing amplitude $r\equiv |\xi| \ne 0$, to a nonzero contribution in $\Delta W$, which also involves two-photon channels, see Eq. \eqref{DeltaW}. To make this argument explicit, we now study the weakly squeezed regime $r\ll 1$. Expanding the squeezing operator to first order in $r$, 
\begin{equation}
S(\xi) = 1 - \frac{r}{2}\left(\textrm{e}^{i\varphi} {a^\dagger}^2 - \textrm{e}^{-i\varphi} {a}^2\right) + \mathcal{O}(r^2), 
\end{equation}
we find (see Appendix  \ref{App1})
\begin{equation}\label{WAlr}
W_{\text{Al}} \underset{r\ll 1}{\simeq} k_{\mathrm{B}} T \ln\left(\frac{1-\textrm{e}^{-\beta \hbar \omega(t_0)}}{1-\textrm{e}^{-\beta \hbar \omega(t_f)}}\right)  + \mathcal{O}(r^2) ,
\end{equation}
so Alicki's work has no linear-squeezing contribution. On the other hand, for the expansion work $W_{\text{exp}}=\tilde{W}_{\text{Al}} - \Delta W$ we obtain $\tilde{W}_{\text{Al}}={W}_{\text{Al}} - \frac{\hbar}{2}[\omega(t_f)-\omega(t_0)]$, with ${W}_{\text{Al}}$ given in Eq.~\eqref{WAlr}, and
\begin{equation}
\Delta W \underset{r\ll 1}{\simeq} - r \hbar \int_{t_0}^{t_f}    \frac{1+\textrm{e}^{-\beta\hbar\omega(t)}}{1-\textrm{e}^{-\beta\hbar\omega(t)}} \cos\left[\varphi-2t\omega(t) \right] \dot{\omega}(t) dt.
\label{DeltaW2}
\end{equation}
Therefore $\Delta W \sim \mathcal{O}(r)$, confirming the inequivalence of Alicki's and expansion work for nonclassical states of light. In view of this inequivalence, we adopt the expansion work as a sound and robust thermodynamic observable to quantify the net mechanical work that can be extracted from the system \cite{tejero:pre24}. Interestingly, it can be shown that the energy change captured by $\Delta W$ and associated with two-photon internal channels does not result in von Neumann entropy changes at order $\mathcal{O}(r)$, and hence can be tentatively classified as work performed on internal degrees of freedom (to maintain squeezing), according to the entropy-based definition of work and heat in quantum thermodynamics \cite{ahmadi:scirep2023, alipour:pra2022}. In any case the expansion work, together with the internal $\Delta W$ and the heat dissipated to the external squeezing bath do fulfill the first law of thermodynamics, providing the necessary thermodynamical consistency.

Since squeezing is a purely quantum effect with no classical counterpart, and squeezed states belong to the so-called \emph{nonclassical states of light}; we refer to the expansion work extracted from squeezing as \emph{nonclassical work}. An interesting conclusion from Eq. (\ref{DeltaW2}) is that the difference $\Delta W$ between Alicki's definition and the expansion work in the weakly squeezed regime can change sign depending on the relation between the squeezing phase $\varphi$ and the expansion protocol as dictated by $\omega(t)$ and $\dot{\omega}(t)$.

\begin{figure}
\centering
\includegraphics[width=.95\linewidth]{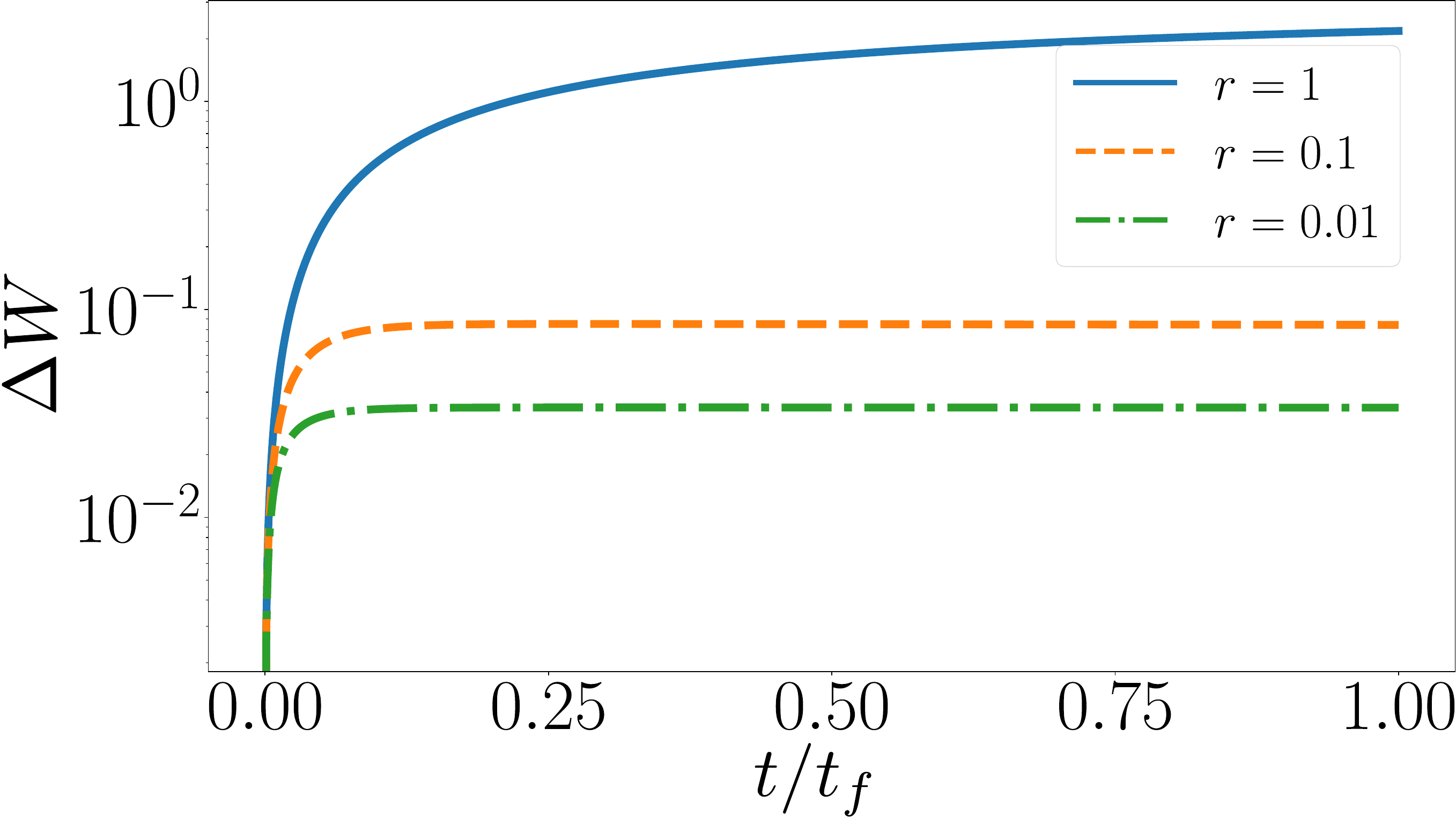}
\caption{Difference between Alicki's work definition and the observed expansion work, $\Delta W (t)= \tilde{W}_{\text{Al}}(t) - W_{\text{exp}}(t)$, resulting from internal two-photon processes in the squeezed system, for different values of the squeezing amplitude $r$ (with $\varphi=0$). The parameters for this expansion protocol are $L_0 = 1$, $v = 2.5 \times 10^{-3}$, $\bar{n}=10$, $t_f = 1000$, $\gamma = 0.01$, and units are such that $\hbar=k_{\mathrm{B}}=1$.}
\label{fig:Wint}
\end{figure}

To test this inequivalence beyond the quasistatic approximation, we now measure numerically both work expressions $\tilde{W}_{\text{Al}}$ and $W_{\text{exp}}$ in finite-time protocols and determine their difference $\Delta W$. In particular, we have performed expansion experiments at constant speed $v$ and $r=0.01, 0.1$ and $1$ by solving numerically Eq. \eqref{eqn:lindblad_squeezing} from an initial length $L_0$ to a final length $L_\text{fin} = L_0 + v t_\text{fin}$, measuring both $\tilde{W}_{\text{Al}}$ and $W_{\text{exp}}$. Determining beforehand the sign of $\Delta W$ in Eq. (\ref{DeltaW}) is not straightforward due to the combination of oscillating components appearing in the integral. Figure \ref{fig:Wint} shows the time evolution of the internal work, showing that $\Delta W(t)>0$ for all $t$. This implies that, for these parameters, Alicki's work overestimates the mechanical work that can be extracted from the system. This is due to the energy utilized in maintaining light squeezing via the two-photon processes mentioned above. These two-photon effects and their dependence with driving parameters will be studied in depth in future works \cite{tejero:inprep}.

From a resource-theoretic perspective \cite{Ng2018,Lostaglio_2019}, squeezing constitutes an athermal (quantum) resource, i.e., a structured deviation from equilibrium that cannot be generated by free thermal operations \cite{narasimhachar2021}. Quantum resources can be quantified through resource monotones, i.e., functions that are nonincreasing under free operations, and can be used in multiple tasks such as sensing and metrology.  In our case, $\Delta W$ can be seen as a \emph{passive energy}, or energy that cannot be extracted because it’s \emph{locked} into preserving the nonclassical resource (squeezing). This prevents the quantum resource from degradation, aligning with the general idea that maintaining quantum correlations costs energy.

\section{A squeezing Otto engine}
We next extend the Otto cycle, a canonical model of heat engine, to nonthermal processes. It is customarily composed of two adiabatic strokes, i.e., a compression and an expansion stroke with no heat exchange, and two isochoric strokes, where the system exchanges heat at constant volume with thermal baths at different temperature \cite{tejero:pre24}. To investigate the role of light squeezing as quantum resource apart from thermal driving, we change now the isochoric thermal strokes to \emph{isochoric squeezing strokes}, meaning that thermal baths will have the same, constant temperature during the whole cycle, but each bath will be characterized by a different squeezing amplitude $r_i$ (with $\varphi_i=0$), see the sketch in Fig.~\ref{fig:sketch}. In this way, the squeezing gradient between the baths will be responsible for the net expansion work performed during the whole cycle. 

To be more precise, the system time evolution during a cycle is described by a superoperator $\Lambda_{\text{cycle}} = \prod_{k=1}^4 \;P_k \Lambda_{k}$, acting on the density matrix for the system state. Each label $k$ corresponds to a different stroke, having all the same finite duration $\tau$. The system is initially prepared in a squeezed thermal state $\rho_{\beta}^{\xi_1}$ with squeezing amplitude $r_1=|\xi_1|$ and an average number of photons $\bar{n}$ that is kept constant throughout the cycle. The first stroke is an adiabatic compression from length $L_1$ to $L_2<L_1$, while the third one is an expansion from $L_2$ to $L_1$. These strokes imply the modification of the cavity frequency $\omega(t)\propto L(t)^{-1}$ from $\omega_1$ to $\omega_2>\omega_1$ in the first cycle and the other way around in the third. During the expansion and compression strokes the system is isolated, so the evolution is unitary for $k=1,3$ so $\Lambda_{k}\rho=U_k\rho U_k^\dagger$, with 
\begin{equation}
    U_{k} = \exp\left[ -\frac{i}{\hbar} \int_{t_k}^{t_{k+1}} H(t) dt \right],
\end{equation}
and $t_k=(k-1) \tau$. We choose the cavity length to vary linearly at constant speed $v$, so $L(t) = L_0 \pm |v|t$ for the compression ($-$, $k=1$) and expansion ($+$, $k=3$) strokes. 
Due to isolation, no heat is exchanged with the baths during these strokes, so all energy variations during expansion and compression can be interpreted in terms of work, most as extractable expansion work $W_{\text{exp}}^{(k)}$ but with some energy $\Delta W^{(k)}$ leaked to maintain squeezing.

The second and fourth strokes are isochoric squeezing processes, meaning that the constant-volume system is put in contact with squeezing thermal baths with squeezing amplitudes $r_2$ (for $k=2$) and $r_1<r_2$ (for $k=4$).  Dynamics during these strokes ($k=2,4$) is thus modeled by a dissipative superoperator 
\begin{equation}
\Lambda_{k} = \exp\left( \int_{t_k}^{t_{k+1}} \mathcal{L}_{k} dt \right),
\end{equation}
with Liouvillian $\mathcal{L}_{k}$
\begin{equation}
   \mathcal{L}_{k} \rho = -\frac{i}{\hbar} \left[H(t_k),\rho\right] + \mathcal{D}_{\bar{n},\xi_k}[\rho],
\end{equation}
where $\mathcal{D}_{\bar{n},\xi_k}$ is the corresponding dissipator defined in Eq. (\ref{eqn:lindblad_squeezing}). In these two strokes, energy exchanges with the baths correspond to heat $Q^{(k)}$, with $k=2,4$. 

\begin{figure*}
\includegraphics[width=0.9\textwidth]{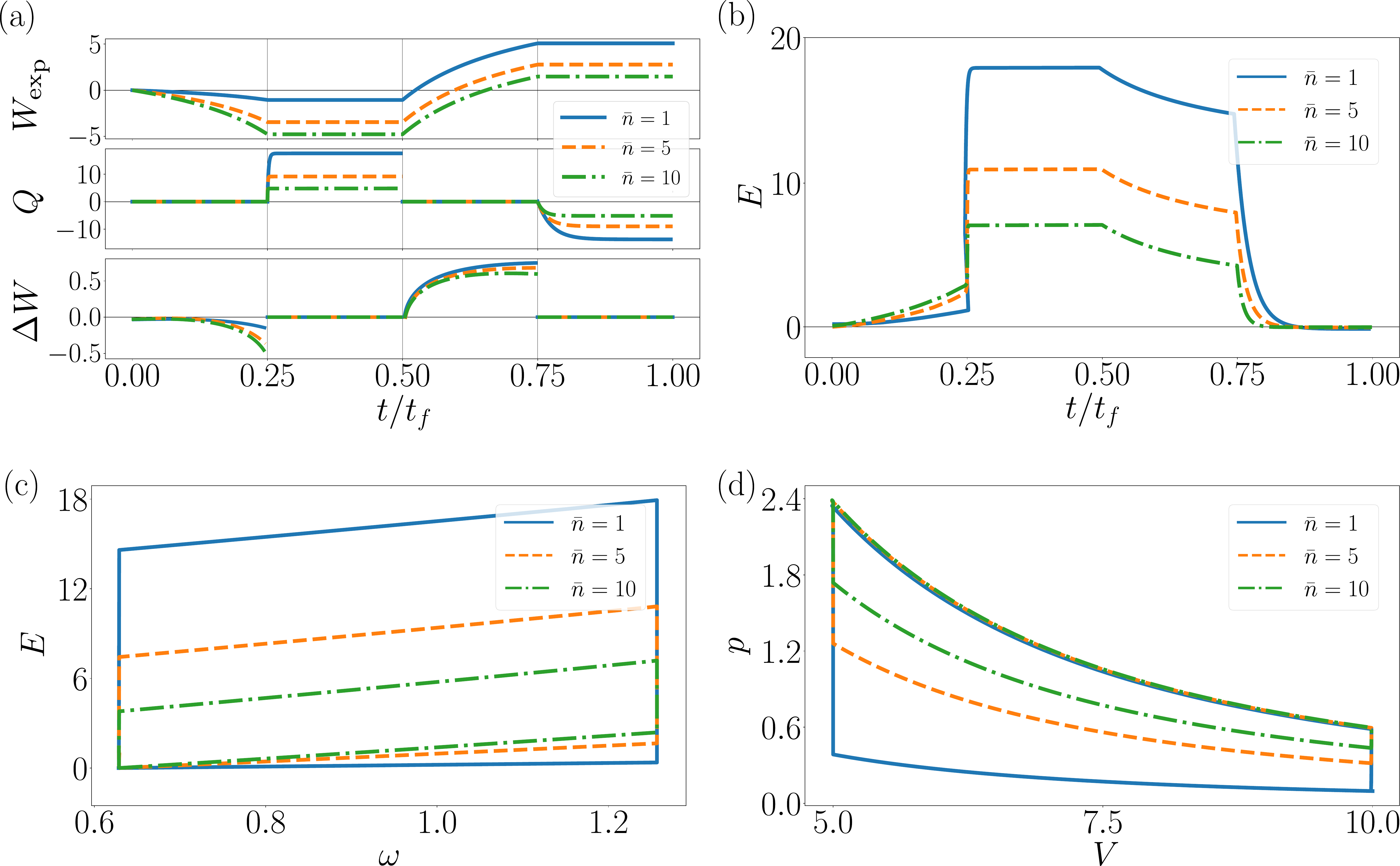}
\caption{(a) Energy balance for a squeezing Otto engine. Top panel shows the expansion work $W_\text{exp}$, middle panel displays the heat $Q$ exchanged with the squeezing baths, and the bottom panel shows the energy $\Delta W$ leaked to internal degrees of freedom to maintain squeezing. Consistency with the first law of thermodynamics is fulfilled in all cases. The squeezing amplitudes are $r_1 = 0.1$ and $r_2 = 10$, for different values of the bath average photon number $\bar{n}$. (b) Time evolution of the total energy during an Otto cycle. Panel (c) shows the energy as a function of the frequency $\omega(t)$ for these cycles, while panel (d) displays the pressure-volume diagrams for the same cases. The parameters used in the cycle simulation are: $\omega_0 = 2\pi, \tau = 1000, \gamma = 0.01, L_0 = 10, |v|=0.005$, and units are such that $\hbar=k_{\mathrm{B}}=1$.}
\label{fig:cycles}
\end{figure*}

At the end of each stroke a projective measurement in the energy eigenbasis, $P_k\rho \equiv \Pi_k\rho\Pi_k$, is performed in order to ensure the possibility of a closed cycle \cite{zheng:pre16,tejero:pre24}. This is modeled by the projector $\Pi_k=\sum_{n=0}^\infty \ket{n}\bra{n}_k$, where $\left\{ \ket{n}\right\}_k$ represents the eigenstates of the Hamiltonian at the end of stroke $k$. In this way, the state after the measurement (time $t_k^+$) with respect to the state before the measurement (time $t_k$) is hence $\rho(t_k^+)=\Pi_k\rho(t_k)\Pi_k$.

\section{Thermodynamical analysis of the squeezing Otto Engine}

To study the performance of the squeezing Otto engine, we solve numerically Eq. \eqref{eqn:lindblad_squeezing} for the strokes previously described. We now show the results for thermal baths with squeezing $r_1 = 0.1$ and $r_2 = 10$, and different average number of photons $\bar{n}$. Figure \ref{fig:cycles}(a) displays the energy balance during a cycle for the squeezing Otto engine, showing the time evolution of the measured expansion work $W_{\text{exp}}(t)$ (top panel), the heat $Q(t)$ exchanged with the squeezing baths during the dissipative strokes (middle panel), and the energy $\Delta W(t)$ leaked to internal degrees of freedom to maintain squeezing during compression and expansion (bottom panel). As expected, consistency with the first law of thermodynamics establishing the conservation of energy is fulfilled for all $t$. 
Indeed, the temporal evolution of the total energy along the cycle is depicted in Fig. \ref{fig:cycles}(b) at the steady state functioning of the engine. Figure \ref{fig:cycles}(c) shows an equivalent plot of energy as a function of the frequency $\omega(t)$ of the cavity mode, while Fig.~\ref{fig:cycles}(d) presents the $p$--$V$ diagram of the cycles. According to Eq.~\eqref{eqn:expansion}, the area enclosed by these $p$--$V$ curves represents the net work extracted, which is notably larger for lower temperatures. In fact, all Figs. \ref{fig:cycles}(a)--(d) reveal a strong dependence of the cycle performance on $\bar{n}$. Although the shape of the curves remains similar, the amount of energy transferred to the system due to the squeezing gradient is larger for lower values of $\bar{n}$. This confirms the phenomenological picture that a quantum resource such as squeezing is more productive in terms of energy when thermal effects are weaker (see, however, below). 

Next, we analyze the heat exchanged during the cycle for completeness. During the isochoric strokes 2 and 4 the system is coupled to squeezing baths characterized by a constant $\bar{n}$ and different squeezing parameters $\xi_k$, see dissipator in the Lindblad Eq. \eqref{eqn:lindblad_squeezing}. No work is performed during these dissipative strokes, and the change of internal energy is solely due to heat transfer, see middle panel in Fig. \ref{fig:cycles}(a). For a squeezed thermal state as the one described by Eq. (\ref{eq:squeezed}), characterized by a squeezing parameter $\xi_k$, the mean photon number is \cite{kim:pra89}
\begin{equation}
    \langle n \rangle_{\xi_k} = \bar{n} + (1 + 2 \bar{n}) \sinh^2 r_k.
\end{equation}
Therefore, assuming that the system is in a squeezed thermal state at the beginning and at the end of the dissipative strokes (a reasonable assumption for long enough stroke durations $\tau$, as the one chosen here $\tau=10^3$), the heat transfer during the $k$th stroke is 
\begin{equation}
Q_k = \hbar \omega(t_k) \left[ \langle n(t_{k+1}) \rangle_{\xi_{k+1}} - \langle n(t_k) \rangle_{\xi_{k}} \right]  .
\label{Qdissip}
\end{equation}
The net heat exchanged then depends directly on the values of the squeezing parameters and frequencies, as $Q_{2} + Q_{4} = \hbar (1 + 2 \bar{n}) (\sinh^2 r_2 - \sinh^2 r_1) (\omega_2 - \omega_1)$. It is positive for $r_2>r_1$ and $\omega_2 > \omega_1$, so it contributes to the energy balance that enables net work extraction, as can be deduced from Fig. \ref{fig:cycles}(b). 

\begin{figure}
\includegraphics[width=0.9\linewidth]{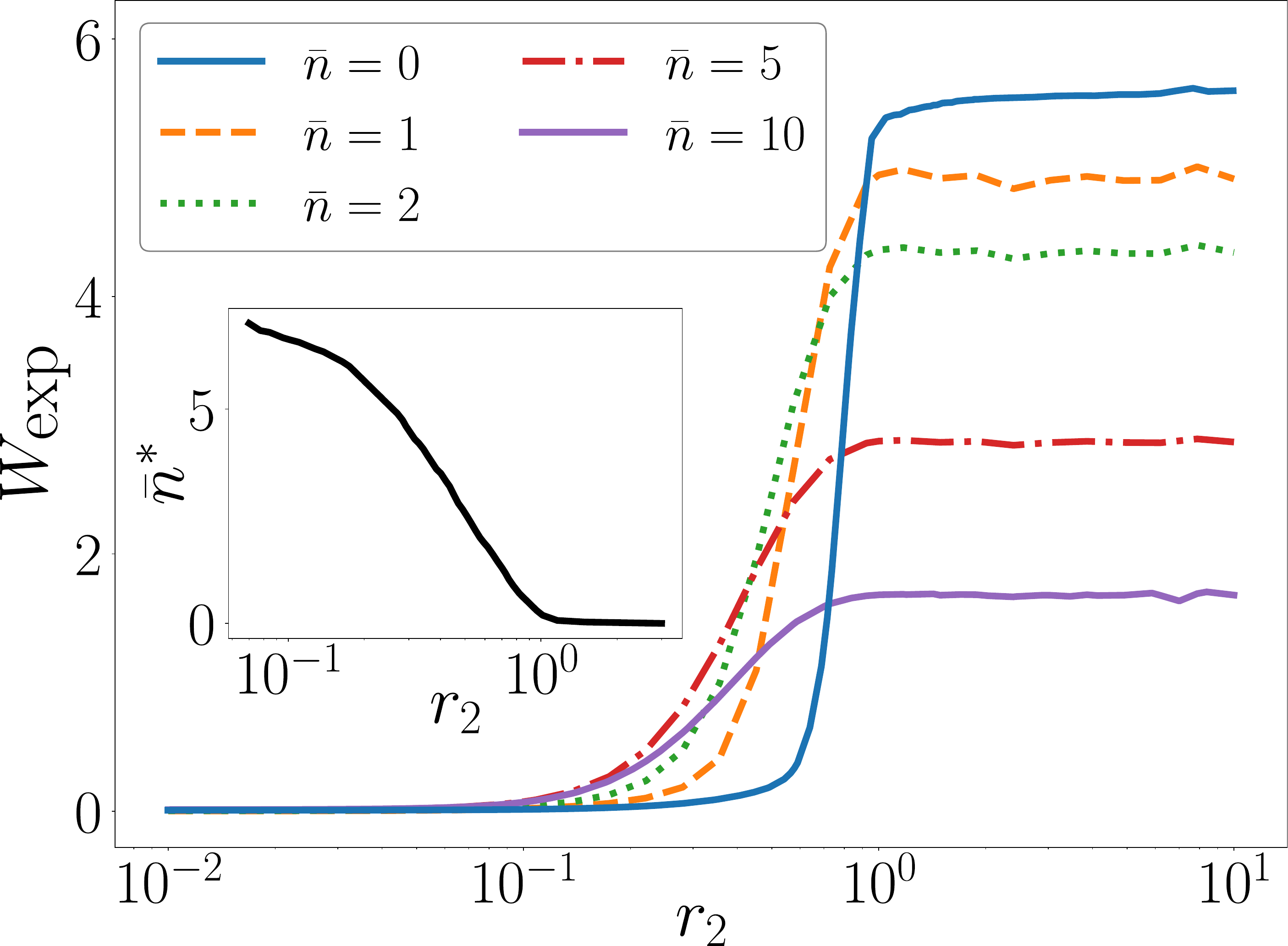}
\caption{Expansion work for an Otto cycle as a function of the high-squeezing amplitude $r_2$, for different values of $\bar{n}$, and fixed $r_1=0.1$. Note the crossing of curves for intermediate squeezing $r_2$, and the large-$r_2$ saturation. This reveals a nontrivial interplay between squeezing and thermal effects which can be harnessed for optimal work output. Inset: Optimal excitation number $\bar{n}^*(r_2)$ maximizing work output as a function of $r_2$. Simulation parameters as in Fig.~\ref{fig:cycles}.}
\label{fig:W_exp_r_n}
\end{figure}

To further understand the interplay between squeezing and thermal effects, we have also considered several cycles with $r_1=0.1$, fixed values of $\bar{n}$ and varying values of the squeezing amplitude $r_2$. Figure~\ref{fig:W_exp_r_n} shows the net expansion work extracted from the cycle as a function of $r_2$ for different values of $\bar{n}$. The expansion work increases monotonically with $r_2$ for each $\bar{n}$, saturating to a maximum expansion work $W_\text{exp}^\text{max}(\bar{n})$ for large enough $r_2$. We find that the asymptotic $W_\text{exp}^\text{max}(\bar{n})$ decreases for increasing $\bar{n}$, as expected from the detrimental effect of thermal fluctuations for quantum work extraction described above. Indeed, a similar temperature-squeezing interplay has also been observed in the charging process of quantum batteries \cite{centrone:2021pra}. Quite remarkably, however, for intermediate values of the squeezing $r_2$, there exists a nontrivial value $\bar{n}^*(r_2)$ which maximizes the expansion work output. This behavior is already apparent from the crossing of the $W_\text{exp}(r_2)$ curves in Fig.~\ref{fig:W_exp_r_n}. Solving the equation $\left.\partial W_{\text{exp}} (r_2, \bar{n}) / \partial \bar{n}\right|_{\bar{n}=\bar{n}^*}=0$, we obtain $\bar{n}^*(r_2)$, see inset to Fig.~\ref{fig:W_exp_r_n}. Interestingly, while $\bar{n}^*(r_2)$ decreases monotonically with $r_2$, for intermediate squeezing it takes a nonzero value, meaning that some degree of thermal activity helps in extracting usable energy from squeezing baths even if there is no temperature gradient. This reveals a nontrivial interplay between squeezing and thermal effects which can be harnessed for optimal work output.

Finally, in order to assess the performance of the squeezing Otto engine, we now discuss its efficiency assuming squeezed thermal states at the beginning and the end of the dissipative strokes, see Eq. \eqref{Qdissip} above. The efficiency $\eta$ is defined as the ratio of the net extractable work output over the heat absorbed. In our case, the net extractable work is given by the expansion work $W_\text{exp}$ integrated over the whole cycle, see Eq. \eqref{eqn:expansion} and Fig.~\ref{fig:cycles}(d). Moreover, stroke 2 acts as the \emph{energy-absorbing} stroke, where the system absorbs heat due to the transition to higher squeezing ($ r_2 > r_1 $), while in stroke 4 the system releases heat in contact to the low-squeezing bath. Therefore, the efficiency of the squeezing engine in extracting mechanical work is simply $\eta \equiv W_{\mathrm{exp}}/Q_{2}$. Now, since the engine operates in a stationary cycle, $\Delta E_{\mathrm{cyc}} = 0 = \tilde{W}_\mathrm{Al} - (Q_2 + Q_4)$, 
see Fig. \ref{fig:cycles}(b), so the first law of thermodynamics imposes that 
\begin{eqnarray}
\tilde{W}_\mathrm{Al}  &=& Q_{2} + Q_{4}  \\
&=& \hbar (1 + 2 \bar{n}) (\sinh^2 r_2 - \sinh^2 r_1) (\omega_2 - \omega_1). \nonumber
\end{eqnarray}
Since $W_{\text{exp}} = \tilde{W}_\mathrm{Al} - \Delta W$ and $\Delta W>0$ in our case, the net extractable work is bounded by Alicki's expression, as seen in Fig \ref{fig:Wint}. Writing now $W_{\mathrm{exp}} = \varepsilon\left({Q_{2} + Q_{4}}\right)$, with $0<\varepsilon<1$ a positive constant defined as $\varepsilon \equiv 1- \Delta W/\tilde{W}_{\text{Al}}$, we find that
\begin{equation}
\eta \equiv \frac{W_{\mathrm{exp}}}{Q_{2}}  = \varepsilon \eta_{\mathrm{Otto}} ,
\end{equation}
where $\eta_{\mathrm{Otto}} = 1 - \omega_1/\omega_2$ is the standard Otto-engine efficiency. Physically, the reduced efficiency for the mechanical work output arises because a fraction of the energy from the high-squeezing bath is used to drive the two-photon internal squeezing process, lowering the net mechanical work extracted compared to the standard case.

\section{Conclusions}
In this work, we have used light squeezing as a quantum resource to generate work via the radiation pressure exerted on an optomechanical mirror. Our analysis has shown that Alicki's definition of work in quantum systems does not capture the net extractable mechanical work, since it misses the energy leaked to the squeezed quantum degrees of freedom via two-photon processes. We have also proposed a quantum Otto engine to extract work from thermal baths that pump squeezing in and out while keeping constant the temperature of the working medium, proving that mechanical work can be extracted from a purely quantum feature, even in the absence of a temperature gradient, although thermal effects can help optimize work output.

Our approach thus clarifies the role of squeezing as an independent thermodynamic resource, decoupled from additional thermal drivings \cite{brandao:prl2013, binder_18,goold:jpa2016, scully:science03, dillenschneider:epl2009, horodecki:natcomm2013, perernau:prx2015, brown:njp2016, huang:pre2012, niedenzu:natcomm2018, manzano_pre16, centrone:2021pra}. 
For instance, Ref. \cite{niedenzu:natcomm2018} demonstrated efficiency enhancements beyond Carnot bounds using both squeezed baths and thermal gradients. Similarly, Ref. \cite{manzano_pre16} analyzed entropy production in squeezed reservoirs under temperature differences. In contrast, our work completely decouples squeezing from any thermal gradient by designing an isothermal Otto engine driven solely by squeezing baths at constant temperature. This isolation clarifies nonclassical advantages such as optimal work extraction at nonzero temperatures, distinct from prior enhancements tied to mixed drivings.

The challenge remains to test these findings in experimental realizations of squeezed quantum engines, at reach now due to the recent technological advances in the field. In particular, several state-of-the-art experimental platforms exist now where squeezed reservoirs are already available. In cavity optomechanics, squeezed vacuum fields generated with nonlinear optical parametric oscillators have been successfully injected into high-finesse cavities to enhance quantum correlations and reduce noise \cite{jabri2022, Croquette2022}. In the microwave domain, superconducting circuits provide an alternative route, where Josephson junctions and traveling-wave amplifiers can be used to produce stable squeezed microwave fields that can be coupled to resonators and electromechanical elements \cite{dassonneville2021, Qiu2022}. Hybrid electro-optomechanical systems further allow the transfer of squeezing between the optical and microwave domains, opening up a versatile arena to explore quantum thermodynamics with engineered reservoirs \cite{Zeng2025}. 
For instance, according to Fig.~\ref{fig:W_exp_r_n} net mechanical work extraction exceeding 10\% of the thermal baseline requires a squeezing parameter in the range $r\in[0.1,1]$, which corresponds to experimental squeezing levels of $S\equiv -10 \log_{10} (\textrm{e}^{-2r}) \sim 0.9-9.0 \text{ dB}$ \cite{walls_2025}. 
Current platforms achieve $\sim 3.0$ dB (e.g., injected at LIGO, scalable to $\sim 6.0 \text{ dB}$ \cite{tse19a}), and $>3.0 \text{ dB}$ stabilized in microwave modes \cite{dassonneville2021}. Thus, $r \sim 1$ ($S \approx 9.0 \text{ dB}$) seems experimentally accessible. Moreover, hybrid cavity platforms can implement optomechanical cycles with cavity lengths $L \sim 10^{-3} - 1 \text{ m}$, since frequencies are $\omega \sim $ GHz \cite{halaski2024}. In this way, these technological platforms pave the way for the experimental demonstration of nonclassical work extraction from squeezing.

\acknowledgments
We acknowledge funding from the Ministry of Science, Innovation and Universities, the Ministry for Digital Transformation and of Civil Service, and AEI (DOI 10.13039/501100011033) of the Spanish Government through Projects No. PID2023-149365NB-I00, No. PID2020-113681GB-I00, No. PID2021-128970OA-I00, No. C-EXP-251-UGR23, No. P20\_00173, No. FPU20/02835 and QUANTUM ENIA project call - Quantum Spain project, and by the European Union through the Recovery, Transformation and Resilience Plan - NextGenerationEU within the framework of the Digital Spain 2026 Agenda, and also the FEDER/Junta de Andaluc\'{\i}a Program No. A.FQM.752.UGR20. We are also grateful for the computing resources and related technical support provided by PROTEUS, the supercomputing center of Institute Carlos I for Theoretical and Computational Physics in Granada, Spain.

\bibliography{phys.bib}

\appendix

\onecolumngrid

\section{Master equation with squeezing}
\label{App1}

The dynamical behavior of an open quantum system, described by a density matrix $\rho(t)$, interacting with a squeezing thermal bath is given by a Lindblad-type Master Equation  \cite{manzano:aip20,rivas_12,gardiner_00} in the form 
\begin{equation}\label{eqn:lindblad}
\dfrac{d \rho(t)}{dt} = - \dfrac{i}{\hbar} \left[H(t),\rho(t) \right] + \mathcal{D}[\rho],
\end{equation}
where $H(t)$ is the Hamiltonian of the system, which may be time dependent, and $\mathcal{D}[\rho]$ is a dissipator modeling the Markovian interaction with the bath. The explicit form of the dissipator is  \cite{gardiner_00}
\begin{eqnarray}\label{eqn:dissipator_squeezing}
    \mathcal{D}[\rho] &=& \gamma (N + 1) \left(L\rho L^{\dagger} - \dfrac{1}{2} \left\{L^{\dagger} L, \rho \right\}\right) 
    + \gamma N \left(L^{\dagger}\rho L - \dfrac{1}{2} \left\{L L^{\dagger}, \rho \right\}\right) \nonumber \\
    &-& \gamma M \left(L^{\dagger}\rho L^{\dagger} - \dfrac{1}{2} \left\{L^{\dagger} L^{\dagger}, \rho \right\}\right) 
    - \gamma M^* \left(L\rho L - \dfrac{1}{2} \left\{L L, \rho \right\}\right),
\end{eqnarray}
where $L$ is a jump operator, $\gamma$ the decay rate controlling the interaction, and $M,N$ are two constants defined as
\begin{eqnarray}\label{eqn:N_M_constants}
    N &\equiv& \bar{n} \left(\cosh^2 \xi + \sinh^2 \xi \right) + \sinh^2 \xi, \nonumber\\
    M &\equiv& -\left(1+2\bar{n}\right)\cosh\xi \sinh \xi ,
\end{eqnarray}
where $\bar{n} = \left(\exp\left[\beta \hbar \omega\right]-1 \right)^{-1}$  is the average number of excitations of the thermal bath for a given temperature $T$ and in resonance with a frequency $\omega$. Finally, $\xi$ represents the squeezing parameter. To recover the general customary Lindblad form, we define a modified squeezed jump operator
\begin{equation}\label{eqn:squeezed_operators}
    \hat{L} := S(\xi)LS(\xi)^{\dagger} = L \cosh\xi + L^{\dagger}\sinh \xi,
\end{equation}
where $S(\xi) = \exp\left[\frac{1}{2}\left(\xi^* a^2 - \xi a^{\dagger^2}  \right)\right]$ is the squeezing operator. Therefore, Eq. (\ref{eqn:lindblad}) can be written as 
\begin{eqnarray}\label{eqn:lindblad_squeezing2}
    \dfrac{d \rho}{dt} = &-& \dfrac{i}{\hbar} \left[H,\rho \right] + \gamma (\bar{n} + 1) \left(\hat{L}\rho \hat{L}^{\dagger} - \dfrac{1}{2} \left\{\hat{L}^{\dagger} \hat{L}, \rho \right\}\right) 
    + \gamma \bar{n} \left(\hat{L}^{\dagger}\rho \hat{L} - \dfrac{1}{2} \left\{\hat{L} \hat{L}^{\dagger}, \rho \right\}\right),
\end{eqnarray}
which resembles the general expression for a Lindblad master equation. For real and small values of the squeezing parameter, $|\xi| = r \ll 1$, the constants $N$ and $M$ can be expanded up to $\mathcal{O}(r^3)$ as
\begin{eqnarray}\label{eqn:N_M_approx}
    N &\simeq& \bar{n} + \left(1+3\bar{n} \right)r^2  , \nonumber\\
    M &\simeq& -\left(1+2\bar{n}\right)r  ,
\end{eqnarray}
giving rise to a simplified version of the squeezed dissipator
\begin{eqnarray}\label{eqn:dissipator_squeezing_expanded}
    \mathcal{D}[\rho] &=& \gamma \left[\bar{n} + 1 + \left(1+3\bar{n} \right)r^2 \right] \left(L\rho L^{\dagger} - \dfrac{1}{2} \left\{L^{\dagger} L, \rho \right\}\right) 
    + \gamma  \left[\bar{n} + \left(1+3\bar{n} \right)r^2 \right] \left(L^{\dagger}\rho L - \dfrac{1}{2} \left\{L L^{\dagger}, \rho \right\}\right) \nonumber \\
    &+& \gamma \left[\left(1+2\bar{n}\right)r \right]\left(L^{\dagger}\rho L^{\dagger} - \dfrac{1}{2} \left\{L^{\dagger} L^{\dagger}, \rho \right\}\right) 
    + \gamma \left[ \left(1+2\bar{n}\right)r \right] \left(L\rho L - \dfrac{1}{2} \left\{L L, \rho \right\}\right) \nonumber\\
    &=& \mathcal{D}_0[\rho] + \gamma \left(1+2\bar{n}\right)r \left[ f\left(L^{\dagger}, L^{\dagger}\right)+f\left(L, L\right) \right] + \gamma\left(1+3\bar{n} \right)r^2 \left[ f\left(L, L^{\dagger}\right)+f\left(L^{\dagger}, L\right) \right],
\end{eqnarray}
where $f$ is an operator function defined as $f(A,B) : = A\rho B - \frac{1}{2}\left\{B A, \rho \right\}$, while $\mathcal{D}_0[\rho]$ is the usual thermal dissipator without squeezing terms.

\section{Analytical calculation of the work for a quasistatic expansion in the weak squeezing limit}
\label{App2}

Consider an open squeezed optomechanical-cavity model described by a quantum harmonic oscillator. The Hamiltonian describing the system is $H(t) = \hbar \omega(t) a^\dagger a$, and the steady state of the squeezed Lindblad-type equation that describes the time evolution of this open system is a squeezed thermal state, $\rho_\beta^{\xi} = S(\xi) \rho_\beta S^\dagger(\xi)$. We now assume that the cavity undergoes a quasistatic expansion, so $\omega(t)$ will decrease in time while the state of the system remains thermal (and squeezed) with the Hamiltonian $H(t)$. To compute analytically the total amount of work in a process using both Alicki's and expansion-work definitions, we expand the squeezing operator $S(\xi)$ up to $\mathcal{O}(r^2)$, with $\xi = r\textrm{e}^{i\varphi}$,
\begin{equation}
S(\xi) = \textrm{e}^{\frac{1}{2}\left(\xi^* a^2 - \xi a^{\dagger^2}  \right)} \underset{r\ll 1}{\simeq} 1 - \frac{r}{2}\left(\textrm{e}^{i\varphi} {a^\dagger}^2 - \textrm{e}^{-i\varphi} {a}^2\right) + \mathcal{O}(r^2).
\label{eq:sexp}
\end{equation}
This implies that the general squeezed thermal state $\rho_\beta^{\xi}(t) = S(\xi) \rho_\beta(t) S^\dagger(\xi)$, with $\rho_\beta(t)=\mathcal{Z}^{-1}\textrm{e}^{-\beta H(t)}$, can be expanded as
\begin{eqnarray}\label{eqn:smsup_rho}
\rho_\beta^{\xi}(t) \underset{r\ll 1}{\simeq} \left[ 1 - \frac{r}{2}\left(\textrm{e}^{i\varphi} {a^\dagger}^2 - \textrm{e}^{-i\varphi} {a}^2\right) \right] \rho_\beta(t) \left[1 - \frac{r}{2}\left(\textrm{e}^{-i\varphi} {a}^2 - \textrm{e}^{i\varphi} {a^\dagger}^2\right) \right]
= \rho_\beta(t) + \frac{r}{2} \left[\rho_\beta(t), \left(\textrm{e}^{i\varphi} {a^\dagger}^2 - \textrm{e}^{-i\varphi} {a}^2\right)\right],
\end{eqnarray}
up to $\mathcal{O}(r^2)$. We now use this expansion to calculate the work performed by a quasistatic expansion using both Alicki's definition and the expansion work done by radiation pressure.

\subsection{Alicki's definition of work}
In this case, work is defined as 
\begin{equation}
W_{\text{Al}} := - \int_{t_0}^{t_f} \text{Tr}\left[ \dot{H}(t) \rho_\beta^{\xi}(t) \right] dt ,
\end{equation}
which, using the explicit expression of the Hamiltonian, gives
\begin{equation}
W_{\text{Al}} = -\hbar \int_{t_0}^{t_f} \dot{\omega}(t) \tr\left(\rho_\beta^{\xi}(t) a^{\dagger} a\right) dt = -\hbar \int_{t_0}^{t_f} \dot{\omega}(t) \sum_{n=0}^\infty n \braket{n|\rho_\beta^{\xi}(t)|n} dt  ,
\label{WAl3}
\end{equation}
where we have expanded the trace in terms of the Fock basis $\{\ket{n}\}_{n=0}^{\infty}$, such that $a\ket{n}=\sqrt{n}\ket{n-1}$ and $a^\dagger\ket{n}=\sqrt{n+1}\ket{n+1}$, with $a,~a^\dagger$ the ladder operators. Note also that, taking conjugates, $\bra{n}a^\dagger=\sqrt{n}\bra{n-1}$ and $\bra{n}a=\sqrt{n+1}\bra{n+1}$. Assuming now a quasistatic expansion for small squeezing so the state is a squeezed thermal state expanded as in Eq. \eqref{eqn:smsup_rho}, we find
\begin{equation}\label{eqn:sm_averagerho_alicki}
\braket{n|\rho_\beta^{\xi}|n} \underset{r\ll 1}{\simeq} \mathcal{Z}^{-1}\braket{n|\textrm{e}^{-\beta\hbar \omega(t) a^\dagger a}|n} + \frac{r}{2} \mathcal{Z}^{-1} \braket{n|\left[\textrm{e}^{-\beta\hbar \omega(t) a^\dagger a}, \left(\textrm{e}^{i\varphi} {a^\dagger}^2 - \textrm{e}^{-i\varphi} {a}^2\right)\right]|n} + \mathcal{O}(r^2) = \frac{\textrm{e}^{-\beta\hbar \omega(t) n}}{\mathcal{Z}} + \mathcal{O}(r^2)  , \nonumber
\end{equation}
where we have used that $\braket{n|[\textrm{e}^{-\beta\hbar \omega(t) a^\dagger a},B]|n}=0$ for any arbitrary operator $B$ since the Hamiltonian is diagonal in the Fock basis. Using this in Eq.~\eqref{WAl3}, and noting that $\mathcal{Z}=(1-\textrm{e}^{-\beta \hbar \omega(t)})^{-1}$, we thus obtain
\begin{equation}
W_\text{Al} \underset{r\ll 1}{\simeq}  -\hbar \int_{t_0}^{t_f} dt \frac{\dot{\omega}(t)}{\textrm{e}^{\beta\hbar \omega(t)}-1} + \mathcal{O}(r^2) = k_{\mathrm{B}} T \ln\left(\frac{1-\textrm{e}^{-\beta \hbar \omega(t_0)}}{1-\textrm{e}^{-\beta \hbar \omega(t_f)}}\right)  + \mathcal{O}(r^2) .
\label{WAl4}
\end{equation}

\subsection{Expansion work}
As we have seen in the main text, the expansion work for our optomechanical system can be written as $W_{\text{exp}}=\tilde{W}_{\text{Al}} - \Delta W$, with $\tilde{W}_{\text{Al}}\equiv -\hbar  \int_{t_0}^{t_f} \dot{\omega}(t) \tr [(a^\dagger a + \frac{1}{2})\rho_\beta^{\xi}(t)] dt$ the Alicki's work including the zero-point energy (which can be renormalized out), and
\begin{equation}
\Delta W \equiv - \frac{\hbar}{2} \int_{t_0}^{t_f} \dot{\omega}(t) \tr [(a^2 \textrm{e}^{-i2t \omega(t)} + {a^\dagger}^2 \textrm{e}^{i2t \omega(t)})\rho_\beta^{\xi}(t)] dt  .
\label{DeltaW3}
\end{equation}
Proceeding as in the previous section, it is easy to show that
\begin{equation}
\tilde{W}_{\text{Al}} \underset{r\ll 1}{\simeq} k_{\mathrm{B}} T \ln\left(\frac{1-\textrm{e}^{-\beta \hbar \omega(t_0)}}{1-\textrm{e}^{-\beta \hbar \omega(t_f)}}\right) - \frac{\hbar}{2}[\omega(t_f)-\omega(t_0)] + \mathcal{O}(r^2)  ,
\end{equation} 
where an extra term appears when compared to Eq.~\eqref{WAl4}. This term comes from the zero-point energy contribution, a constant that can be renormalized without affecting our results.

On the other hand, to evaluate Eq.~\eqref{DeltaW3} in a weakly-squeezed thermal state $\rho_\beta^{\xi}(t)$ expanded as in Eq.~\eqref{eqn:smsup_rho}, we note that many of the terms that appear after expanding the resulting product are zero, i.e., $\tr[a^2 \rho_\beta(t)]=\tr[{a^\dagger}^2 \rho_\beta(t)]=\tr[a^2 \rho_\beta(t) a^2]=\tr[{a^\dagger}^2 \rho_\beta(t) {a^\dagger}^2]=\tr[a^4 \rho_\beta(t)]=\tr[{a^\dagger}^4 \rho_\beta(t)]=0$. The only remaining terms in the trace within the integral are
\begin{eqnarray}
\frac{r}{2} \textrm{e}^{i [\varphi-2 t \omega(t)]} \left\{\tr[a^2 \rho_\beta(t) {a^\dagger}^2] - \tr[a^2 {a^\dagger}^2 \rho_\beta(t) ]\right\} &=& \frac{r}{2\mathcal{Z}} \textrm{e}^{i [\varphi-2 t \omega(t)]} \sum_{n=0}^\infty (n+1)(n+2) \left(\textrm{e}^{-\beta\hbar \omega(t) (n+2)} -  \textrm{e}^{-\beta\hbar \omega(t) n}\right) , \nonumber \\
\frac{r}{2} \textrm{e}^{-i [\varphi-2 t \omega(t)]} \left\{\tr[{a^\dagger}^2 a^2  \rho_\beta(t)] - \tr[{a^\dagger}^2 \rho_\beta(t) a^2]\right\} &=& \frac{r}{2\mathcal{Z}} \textrm{e}^{-i [\varphi-2 t \omega(t)]} \sum_{n=0}^\infty n (n-1) \left(\textrm{e}^{-\beta\hbar \omega(t) n} -  \textrm{e}^{-\beta\hbar \omega(t) (n-2)}\right) ,  
\end{eqnarray}
which involve the two-photon excitation/decay processes mentioned in the main text, which are responsible for the nontrivial difference between Alicki's definition and expansion work. Noting that in the second equation above the first two terms, $n=1, 2$, do not contribute to the sum, and making a shift $n \to n+2$ in this second sum, we can put together all terms in the trace to obtain
\begin{equation}
\Delta W \underset{r\ll 1}{\simeq} - r \frac{\hbar}{2} \int_{t_0}^{t_f} dt \dot{\omega}(t) \cos\left[\varphi-2t\omega(t) \right] \left(1-\textrm{e}^{-\beta\hbar\omega(t)} \right) \sum_{n=0}^\infty (n+1)(n+2) \left(\textrm{e}^{-\beta\hbar \omega(t) (n+2)} -  \textrm{e}^{-\beta\hbar \omega(t) n}\right) .
\label{DeltaW4}
\end{equation}
Performing the sums using properties of the geometric series, we finally arrive at
\begin{equation}
\Delta W \underset{r\ll 1}{\simeq} - r \hbar \int_{t_0}^{t_f}    \frac{1+\textrm{e}^{-\beta\hbar\omega(t)}}{1-\textrm{e}^{-\beta\hbar\omega(t)}} \cos\left[\varphi-2t\omega(t) \right] \dot{\omega}(t) dt.
\label{DeltaW5}
\end{equation}

\end{document}